\documentclass[fleqn,usenatbib]{mnras}
\usepackage{newtxtext,newtxmath}
\usepackage[T1]{fontenc}
\DeclareRobustCommand{\VAN}[3]{#2}
\let\VANthebibliography\thebibliography
\def\thebibliography{\DeclareRobustCommand{\VAN}[3]{##3}\VANthebibliography}

\usepackage{graphicx}	
\usepackage{amsmath}	
\usepackage{aas_macros}

\title[Probing AGN with spectropolarimetry...]{Probing AGN with spectropolarimetry: accretion disk and SMBH parameters}

\author[M.Yu. Piotrovich et al.]{
M.Yu. Piotrovich$^1$ \thanks{E-mail: mpiotrovich@mail.ru},
E.S. Shablovinskaya$^{2,3}$,
E.A. Malygin$^2$,
S.D. Buliga$^1$
and T.M. Natsvlishvili$^1$,
\\
$^1$ Central Astronomical Observatory at Pulkovo, 196140, Saint-Petersburg, Russia\\
$^2$ Special Astrophysical Observatory of Russian Academy of Sciences (SAO RAS), 369167, Nizhnij Arkhyz, Zelenchukskiy region, Karachai-Cherkessian\\ Republic, Russia\\
$^3$ Instituto de Estudios Astrofísicos, Facultad de Ingeniería y Ciencias, Universidad Diego Portales, Santiago, Región Metropolitana, 8370191 Chile\\
\\ {\it Accepted for publication in MNRAS.}
}

\pubyear{2023}

\begin{document}
\label{firstpage}
\pagerange{\pageref{firstpage}--\pageref{lastpage}}
\maketitle

\begin{abstract}
The interaction of a supermassive black hole with the matter of an accretion disk in the presence of a magnetic field is the key mechanism of energy release in active galactic nuclei. However, determining the physical parameters of this system, such as the spin and mass of the black hole, the shape and parameters of the rotation of the accretion disk, and the geometry of the magnetic field in the accretion disk is a complex and not completely solved problem. We have previously shown, based on our numerical models, that these estimates can be obtained from just three parameters: the black hole mass, bolometric luminosity, and optical polarization. In this paper, we estimate the accretion disk and black hole parameters for a sample of 14 type 1 Seyfert galaxies. Using the spectropolarimetric data obtained by us, we selected only those objects in which the polarization of optical radiation is generated mainly by the mechanism in the accretion disk. Despite the small statistics, our results for such a sample are consistent with our previous conclusions and show a discrepancy between the disk magnetic field parameters and the classical Shakura-Sunyaev disk model.
\end{abstract}


\section{Introduction}

Rotating supermassive black holes are unique powerful energy machines responsible for the physical processes occurring on the giant scale $\sim 10^{22} - 10^{24}$ cm in galaxies. A large number of AGNs exhibit highly collimated ejections of matter (jets) moving at relativistic velocities in a direction perpendicular to the accretion disk. The size of the jets reaches tens of kiloparsecs, which exceeds the size of some galaxies. The main objects of the AGN are the central supermassive black hole and the accretion disk around the central energy machine. The accretion disk has a complex structure, and quite a lot of theoretical papers have been devoted to describing its structure, starting with the classic work by Shakura and Sunyaev \citep{shakura73,lovelace76,blandford76,camenzind86,heyvaerts89,camenzind90,takahashi90,chiueh91,pelletier92,appl92,bogovalov92,beskin93,eichler93,lery99,lyubarsky09}. However, it has not yet been possible to construct a convincing model of the central machine in AGN, which would lead to the generation of an effective outflow of accreting plasma, i.e. to the formation of a relativistic jet. At present, such an outflow is associated with the presence of a regular magnetic field in the AGN structure, the generation mechanism of which also remains unclear. As is known, a black hole itself is not able to have its own magnetic field, so the generation of a large-scale magnetic field can occur in the accretion disk itself, including the region of the innermost stable circular orbit \citep{blandford77,blandford82,meier99,garofalo10,tchekhovskoy10,daly11,yuan14}. Therefore, the determination of such basic SMBH parameters as spin (dimensionless angular momentum) and magnetic field is the important task of modern astrophysics.

Observations make the main contribution to determining the structure of the accretion disk and the physical parameters of the plasma in it. The study of AGN accretion disks is carried out in the optical and UV bands \citep[e.g., disk size estimation by reverberation mapping,][]{ad_rm}; space X-ray missions are aimed at studying the compact hot corona associated with the disk \citep[e.g., the recent observation of the X-ray polarimeter {\it IXPE},][]{ixpe}. Being sensitive to the anisotropy of the distribution of matter, polarization plays an important role in the study of the central optically unresolvable regions of AGN, including the accretion disk. It is a typical example of a radiative region with a non-spherically symmetric electron density distribution, due to which, as a result of scattering by plasma electrons, the disk radiation becomes polarized. Since the matter of the disk is assumed to be magnetized, the Faraday rotation effect will be observed over the free path of photons in the process of scattering by electrons, and the magnitudes of the degree and position angle of polarization, as well as their dependence on the wavelength, will be completely determined by the geometry of the magnetic field in the radiation region. The method developed by \citet{gnedin84,dolginov95,gnedin97,silantev09} makes it possible to estimate the magnitude and geometry of the magnetic field by analyzing optical polarization. We have previously applied this method to observational data \citep{piotrovich21}. However, at that time we used a heterogeneous sample of type 1 AGNs, where, as is known, the polarization of the continuum can be generated by different mechanisms \citep[see, e.g. the review by][]{Shab20}. In this work, we analyzed the spectropolarimetric data of 14 AGNs with broad emission lines, for which the parameters necessary for our calculations (their bolometric luminosity and the mass of the central SMBH) were known. We have selected only those objects where there are no signs of the contribution of other scattering mechanisms to the observed polarization, forming thus a more homogeneous AGN sample.

\section{Observations and data processing}

\subsection{Observation technique}

In the period from 2017 to 2021, within the authors' proposal together with V.L. Afanasiev (SAO RAS), we observed the AGN sample with the BTA-6m telescope of the SAO RAS. The observational data were obtained using the focal reducer SCORPIO-2 \citep{afanasiev11b} in the spectropolarimetric mode.

To obtain spectropolarimetric data, the input beam is passed through a slit with 2$''$ width for all observations. The slit image is decomposed using a double Wollaston prism into four directions of oscillation of the electric vector 0$^\circ$, 90$^\circ$, 45$^\circ$ and 135$^\circ$. Then, each slit image hits a dispersive element -- a volume phase holographic grating (VPHG). Depending on the redshift $z$ of the object, we used VPHG940@600 and VPHG1026@735\footnote{See \url{sao.ru/hq/lsfvo/devices/scorpio-2/grisms_eng.html}.} with broad filters suppressing the second order of the gratings. The reciprocal dispersion is 2~\AA/pixel and 1.5~\AA/pixel, respectively. CCD~E2V42-90 was used as a detector. The exposure time for each object was selected based on its brightness and weather conditions and ranged from 1 to 2 hours. Immediately after the observations of each object, a standard set of calibration frames is taken: a flat field, a line spectrum of a He-Ne-Ar lamp, and continuum spectra of a 3-dot mask to correct geometric distortions. Also, zero and high polarization standards were taken during the night.

\begin{table*}
\caption {Observation log of a sample of objects. Names of objects, coordinates, magnitude in the $V$ band, type of active galaxy, cosmological redshift, and date of observations are given.} \label{log}
\begin{tabular}{lcccccc}
\hline
Object & RA(2000) & DEC(2000) & Magn. & Type & $z$ & Date \\
\hline
Mrk 957        & 00 41 53.43 & +40 21 17.65 & 15.14 & Sy1   & 0.072931 & 2019/11/23  \\
PG 0052+251    & 00 54 52.12 & +25 25 38.98 & 15.42 & Sy1.2 & 0.155    & 2019/11/24  \\
LEDA 3095506   & 01 09 39.01 & +00 59 50.36 & 17.97 & Sy1   & 0.09264  & 2021/09/10  \\
Mrk 359        & 01 27 32.52 & +19 10 43.84 & 14.22 & Sy1.5 & 0.0168   & 2019/11/23  \\
PG 0137+06     & 01 39 55.77 & +06 19 22.52 & 17.00 & Sy1   & 0.396    & 2021/09/09 \\
Mrk 1018       & 02 06 15.99 & -00 17 29.22 & 15.50 & Sy1.9 & 0.04296  & 2019/11/26 \\
Mkn 590        & 02 14 33.56 & -00 46 00.18 & 13.81 & Sy1.2 & 0.02609  & 2019/11/26 \\
LEDA 11567     & 03 04 17.78 & +00 28 27.28 & 17.38 & NLS1  & 0.04445  & 2019/12/18 \\
LEDA 3095636   & 03 22 13.90 & +00 55 13.44 & 16.57 & Sy1   & 0.18494  & 2021/09/08  \\
QSO B0624+6907 & 06 30 02.50 & +69 05 03.91 & 14.16 & QSO   & 0.374    & 2017/01/30  \\
LEDA 2196699   & 07 58 19.69 & +42 19 35.12 & 17.20 & Sy1   & 0.211224 & 2019/11/25 \\
Mrk 142        & 10 25 31.28 & +51 40 34.88 & 16.12 & Sy1   & 0.04459  & 2017/01/30 \\
PG 1307+085    & 13 09 47.00 & +08 19 48.21 & 15.89 & Sy1.2 & 0.154    & 2017/01/31  \\
LEDA 3096673   & 22 19 18.53 & +12 07 53.20 & 17.19 & Sy1   & 0.08128  & 2021/09/09 \\
\hline
\end{tabular}
\end{table*}

\subsection{Data processing}

Since polarimetric observations of weak sources are extremely sensitive to weather conditions, in 2017-2021 we accumulated observations for only 14 objects, the observation log is given in Table \ref{log}. Data reduction was carried out using the original method, the principle of which is described in \citep{afanasiev12}. As a result, for each object, four spectra are obtained as intensity values versus wavelength: $I_{0}(\lambda)$, $I_{90}(\lambda)$, $I_{45}(\lambda)$ and $ I_{135}(\lambda)$. The Stokes parameters are calculated as:
\begin{equation}
Q'(\lambda) =\frac{ I_{0}(\lambda) - I_{90}(\lambda)D_{Q}(\lambda) }{ I_{0}(\lambda) + I_{90} (\lambda)D_{Q}(\lambda) },
\end{equation}

\begin{equation}
U'(\lambda) =\frac{ I_{45}(\lambda) - I_{135}(\lambda)D_{U}(\lambda) }{ I_{45}(\lambda) + I_{135} (\lambda)D_{U}(\lambda)},
\end{equation}
where $D_Q$ and $D_U$ are the transmission coefficients of the polarization channels, determined from observations of the zero polarization standard, $Q'$ and $U'$ are the Stokes parameters for describing linear polarization in the instrumental reference frame. The transformation of $Q'$ and $U'$ to the celestial reference system is done by multiplying the polarization vector by the rotation matrix:
\begin{equation}
\begin{pmatrix}
Q\\
U
\end{pmatrix}
=
\begin{bmatrix}
\cos(-2\cdot{\rm PA}) & \sin(-2\cdot{\rm PA})\\
-\sin(-2\cdot{\rm PA}) & \cos(-2\cdot{\rm PA})
\end{bmatrix}
\times
\begin{pmatrix}
Q'\\
U'
\end{pmatrix},
\end{equation}
where ${\rm PA}$ is the position angle of the instrument during the observation of the object. The following formulas were used to calculate the degree $P$ and the polarization angle $\varphi$:
\begin{equation}
P(\lambda)= \sqrt{Q^{2}(\lambda)+U^{2}(\lambda)},
\end{equation}

\begin{equation}
\varphi(\lambda) =0.5 \cdot \arctan[U(\lambda) / Q(\lambda) ].
\end{equation}
In the case when $Q$ and $U$ are close to zero and have large measurement errors and, consequently, $P/dP \lesssim 0.5$, where $dP$ is the measurement error of the degree of polarization, we used the unbiased estimate of $P$ \citep{simmons1985}:
\begin{equation}
P_{\rm unbiased} = P \cdot \sqrt{1 - (1.41\cdot dP/P)^{2}}.
\end{equation}
In the observed sample, the unbiased estimate was calculated for three objects -- PG~0052+251, LEDA~3095636 and LEDA~2196699.

\subsection{ISM polarization}

It is known that the observed AGN polarization is a vector sum of the polarization of the source and the polarization of the medium through which the radiation passes. The main additional source of polarization is the interstellar medium (ISM), and its value grows while approaching the Galactic plane and is heterogeneous over the sky. Accounting for the ISM polarization is important because its value can be much larger than the AGN intrinsic polarization.

To assess the influence of ISM effects on our data, we used the \citep{heiles} catalogue data and the \citep{fosalba1,fosalba2} statistical analysis based on it. For objects at galactic latitudes $|b| > 40^\circ$ the observed degree of ISM polarization $P_{\rm ISM} < 0.3$\%, which is less than the error of our observations. From our sample, five objects are located at lower galactic latitudes: Mrk~957, PG~0052+251, QSO~B0624+6907, LEDA~2196699, and LEDA~3096673. Using the \citep{heiles} catalogue, the maximum ISM polarization for stars was determined within $\pm$5$^\circ$ by galactic latitude and longitude from sources. For PG 0052+251 the maximum value of $P_{\rm ISM} = 0.2$\%, and for Mrk~957, QSO~B0624+6907 and LEDA~2196699 $P_{\rm ISM} = 0.5$\%. At the same time, for all these objects, the average value of $Q_{\rm ISM}$ and $U_{\rm ISM}$ stars did not exceed 0.1\%, which allows us to say that the ISM polarization is within the observation errors. The largest value is $P_{\rm ISM} = 0.7$\% for LEDA~3096673. For it, the average value for nearby stars is $Q_{\rm ISM} = -0.2$\% and $U_{\rm ISM} = -0.1$\%. This gives a correction in the observed degree and angle of polarization in $\sim$0.2\% and $\sim$1$^\circ$, respectively, which is again less than the measurement error. In order not to increase the uncertainty, in the case of LEDA~3096673 we will also assume that the ISM effect is negligible.

\begin{figure*}
\includegraphics[bb= 30 110 600 685, clip, scale=0.7]{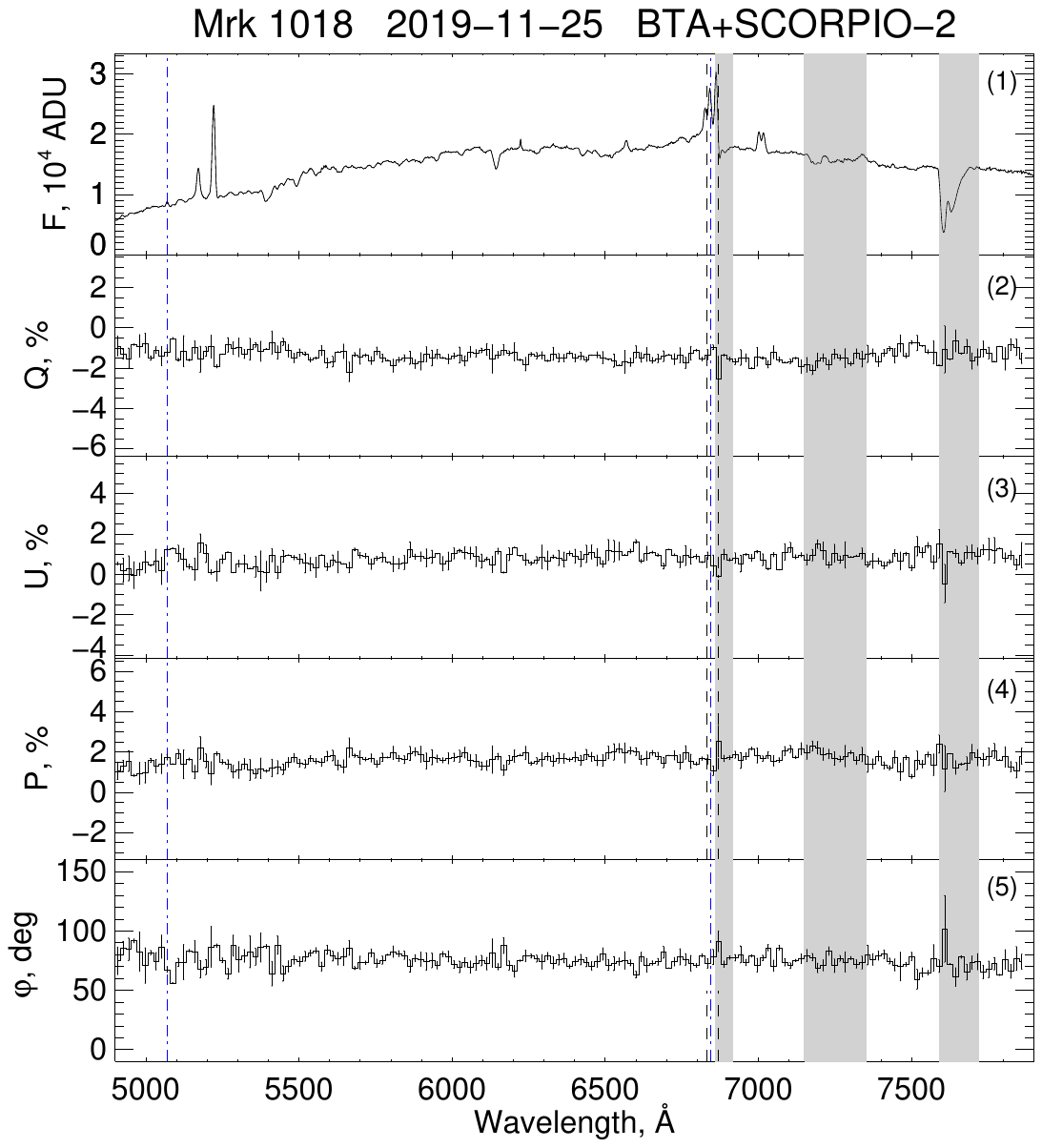}
\caption{Results of processed observations of the object Mrk~1018: integral flux $F$ in ADU units, not corrected for the CCD sensitivity curve (1), Stokes parameters $Q$ (2) and $U$ (3) in per cent, degree of polarization $P$ in per cent (4) and polarization angle $\varphi$ in degrees (5). The polarization parameters are binned in the 18\AA{} window. The vertical blue dash-dotted lines indicate the position (from left to right) of the H$\beta$ and H$\alpha$ lines. The black dashed lines indicate narrow [NII] lines. Atmospheric absorption bands are marked with grey bands.}
  \label{figMrk1018}
\end{figure*}

\section{Observation results}

As a result of observations, we obtained for 14 sample objects the values $Q(\lambda)$, $U(\lambda)$, $P(\lambda)$ and $\varphi(\lambda)$ corrected for instrumental effects, atmospheric depolarization and rotated to the celestial reference system. An example of the resulting spectrum is given in Fig. \ref{figMrk1018}. The results of polarimetric measurements are given in Table \ref{sum}.

\begin{table*}
\caption{Results of polarimetric observations of the AGN sample. Here are given: names of objects; the spectral range of observations; average values of the degree and angle of polarization; the presence (Y) or absence (N) of the dependence of the polarization on wavelength, as well as the presence of signs of equatorial scattering (ES).} \label{sum}
\begin{tabular}{lcccc}
\hline
Object         & Range, \AA & $P$, \%                                  & $\varphi$, $^\circ$                 & $\overrightarrow{P}(\lambda)$ \\ \hline
Mrk 957        & 4900-7900  & \textless{}3.00\textgreater{} $\pm$ 0.92 & \textless{}67\textgreater{} $\pm$ 9 & Y \\
PG 0052+251    & 6000-8200  & 0.44 $\pm$ 0.34                          & 89 $\pm$ 40                         & N \\
LEDA 3095506   & 6000-8000  & 2.45 $\pm$ 0.71                          & 102 $\pm$ 8                         & N \\
Mrk 359        & 4900-7900  & \textless{}2.62\textgreater{} $\pm$ 0.63 & <132> $\pm$ 7                       & Y \\
PG 0137+06     & 6000-8000  & 1.81 $\pm$ 0.50                          & 133 $\pm$ 8                         & N \\
Mrk 1018       & 4900-7900  & 1.62 $\pm$ 0.30                          & 75 $\pm$ 5                          & N \\
Mkn 590        & 4900-7900  & 1.25 $\pm$ 0.29                          & 49 $\pm$ 9                          & N \\
LEDA 11567     & 6000-8000  & 2.58 $\pm$ 0.85                          & 3 $\pm$ 12                          & N \\
LEDA 3095636   & 4900-7900  & <0.74> $\pm$ 0.63                        & 98 $\pm$ 27                         & Y/ES \\
QSO B0624+6907 & 6000-8000  & 1.39 $\pm$ 0.21                          & 139 $\pm$ 8                         & N/ES \\
LEDA 2196699   & 6000-8000  & 0.48 $\pm$ 0.36                          & 98 $\pm$ 44                         & N \\
Mrk 142        & 6000-8000  & 1.29 $\pm$ 0.42                          & 6 $\pm$ 12                          & N \\
PG 1307+085    & 6000-8000  & 1.54 $\pm$ 0.66                          & 119 $\pm$ 13                        & N \\
LEDA 3096673   & 4900-7500  & 2.01 $\pm$ 0.33                          & 116 $\pm$ 6                         & N \\ \hline
\end{tabular}
\end{table*}

For each object, the spectral range of observations was selected so that a broad H$\alpha$ line was present in the resulting spectrum (in cases of PG~0137+06 and QSO~B0624+6907 H$\alpha$ turned out to be out of range, then we used the broad H$\beta$ line). In the case of type 1 AGN, spectropolarimetric observations are necessary not only to analyze the continuum polarization as a function of wavelength but also to check for the presence of specific polarization markers in broad line profiles. It is known that in a number of type 1 AGNs, polarized light is dominated by equatorial scattering when polarization is generated upon reflection of radiation of the AGN central parts from the inner boundary of the torus \citep{smith02,smith05}. This effect generates characteristic features in the profile of broad lines (e.g., the S-shaped profile of the angle $\varphi$) that differ from the polarization of the continuum. Although the analysis of the polarization of AGNs with equatorial scattering makes it possible to independently estimate the parameters of the AGN central parsec \citep{af14,afanasiev15,afanasiev19,sha22}, in this work we analyze data only from those objects where the polarization is generated in the accretion disk. We found signs of equatorial scattering in the objects LEDA~3095636 and QSO~B0624+6907 (marked "ES"{} in Table \ref{sum}). We also note that for a number of objects -- Mrk~957, Mrk~359 and Mrk~142 -- in the H$\alpha$ profiles, we observe slight differences in polarization from the neighbouring continuum. These features have low contrast and are comparable to noise, therefore, in the absence of data with a higher polarization accuracy, we will assume that the polarization in the line is identical to the continuum polarization.

In Table \ref{sum} we give the mean values of the $P$ and $\varphi$ of the continuum polarization for the observed objects. The averaging of polarization degree and angle is carried out across the entire spectral range in which observations have been obtained (see column~2 in Table \ref{sum}). For most of them, we do not observe, within measurement errors, the dependence of $P$ and/or $\varphi$ on the wavelength, except for the following cases.
\begin{itemize}
     \item In the spectrum of Mrk~957, the degree and angle of polarization show dependence on wavelength\footnote{To estimate the values in the $V$ band, we chose the wavelength range 5400-5800\AA, and for the $R$ band, 6400-6800\AA. }: $P(V)=(4.2\pm0.5)$\%, $\varphi(V)=(59\pm4)^\circ$ and $P(R)=(2.3\pm0.4)$ \%, $\varphi(R)=(66\pm4)^\circ$.
     \item A similar dependence is observed in the spectrum of Mrk~359: $P(V)=(3.4\pm0.4)$\%, $\varphi(V)=(125\pm2)^\circ$ and $P(R)= (2.1\pm0.4)$\%, $\varphi(R)=(130\pm4)^\circ$.
     \item In the case of LEDA~3095636 $P(V)=(2.0\pm0.6)$\%, and $P(R)=(0.8\pm0.3)$\%, the angle does not change along the wavelength.
\end{itemize}
We note that the models we use below do not predict such a dependence of the polarization parameters if of the accretion disk origin on wavelengths. Further, we use the values of the polarization averaged along the wavelength to estimate the magnetic fields. However, we note that the AGN polarization in AGN can also have a different nature [for example, polar scattering \citep{smith04}, which is characterized by the dependence $P(\lambda)$ \citep{marin12}].

\section{Methods for determining the SMBH spin}

Obtaining the spin (dimensionless angular momentum) value $a = c J / \left(G M_\text{BH}^2\right)$ (where $J$ is the angular momentum, $M_\text{BH}$ is the black hole mass, $c$ is the speed of light) of the SMBH located at the centre of the AGN is an important problem in modern astrophysics. Spin is believed to play a key role in the generation of relativistic jets in AGN; therefore, it is the power of the relativistic jet that is most often used to determine the spin of the SMBH \citep{daly11}. As a rule, the kinetic power of a relativistic jet is obtained by estimating the magnetic field strength near the SMBH event horizon using several generation mechanisms, such as the Blandford–Znajek \citep{blandford77} mechanism, the Blandford–Payne mechanism \citep{blandford82}, and the Garofalo mechanism \citep{garofalo10}.

\begin{figure}
\includegraphics[bb= 55 35 700 535, clip, width= \columnwidth]{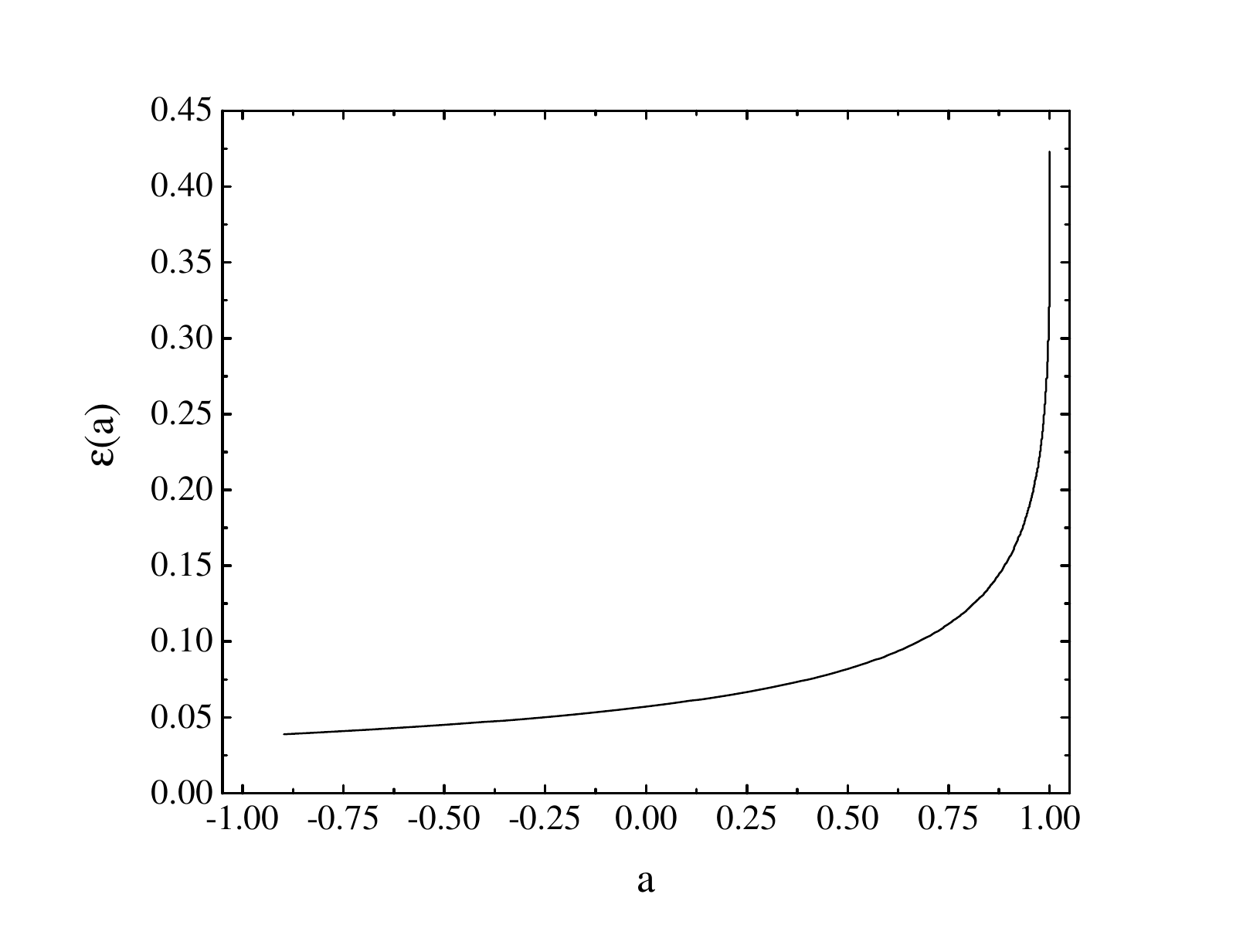}
\caption{Dependence of the radiative efficiency $\varepsilon$ on the spin value $a$.}
  \label{epsilon_fig}
\end{figure}

One of the effective methods for estimating the spin $a$ is the determination of the radiative efficiency $\varepsilon(a)$ of the accretion disk, which essentially depends on the value of the black hole spin \citep{bardeen72,novikov73,krolik07,krolik07b} (see Figure \ref{epsilon_fig}). The radiative efficiency is defined as $\varepsilon = L_\text{bol} / (\dot{M} c^2)$, where $L_\text{bol}$ is the AGN bolometric luminosity and $\dot{M}$ is the accretion rate. In this case, the value of the radiation efficiency should lie within $0.039 < \varepsilon < 0.324$, and the spin $-1.0 < a \leq 0.998$ \citep{thorne74}. The negative value of the spin corresponds to the so-called ''retrograde'' rotation, in which the SMBH and the accretion disk rotate in opposite directions.

There are several models \citep{davis11,raimundo11,du14,trakhtenbrot14,lawther17} relating the radiation efficiency coefficient to such AGN parameters as the SMBH mass $M_\text{BH}$, the angle between the line of sight and the normal to the accretion disk plane $i$, and the bolometric luminosity $L_\text{bol}$, which can be estimated from observations. All these methods use a statistical analysis of AGN observational data and the Shakura-Sunyaev \citep{shakura73} accretion disk model to obtain the relation between the parameters. The \citet{davis11} and \citet{raimundo11} methods are basically the same. The \citet{lawther17} method is a reformulation of the \citet{raimundo11} method. Thus, we decided to use the following three models, which differ from each other to a sufficient extent (we have changed the form of the equations from the original papers for greater uniformity):

\begin{enumerate}
\item \citet{du14}:\\ $\varepsilon \left( a \right) =  0.105 \left(\frac{L_\text{bol}}{10^{46}\text{erg/s}}\right) \left(\frac{L_{5100}}{10^{45}\text{erg/s}} \right)^{-1.5}\times \\ \times M_8 \mu^{1.5}$.

\item \citet{raimundo11}:\\$\varepsilon \left( a \right) = 0.063\left( {\frac{L_\text{bol} }{10^{46}\text{erg/s}}} \right)^{0.99}\times \\ \times\left( {\frac{L_\text{opt} }{10^{45}\text{erg/s}}} \right)^{-1.5} M_8^{0.89} \mu^{1.5}$.

\item \citet{trakhtenbrot14}:\\$\varepsilon \left( a \right) =  0.073 \left(\frac{L_\text{bol}}{10^{46}\text{erg/s}}\right) \left(\frac{\lambda L_\lambda}{10^{45}\text{erg/s}} \right)^{-1.5}\times \\ \times \left(\frac{\lambda}{5100\text{\AA}}\right)^{-2} M_8 \mu^{1.5}\\ \lambda L_\lambda = L_{\rm opt}, \lambda = 4400\text{\AA}$.
\end{enumerate}

\noindent Here $L_{5100}$ is the luminosity at 5100\AA, $M_8 = M_\text{BH} / (10^8 M_{\odot})$ and $\mu = \cos{(i)}$.  For the model from \cite{du14}, we used the Eddington relation $l_\text{E} = L_\text{bol} / L_\text{Edd}$, where $L_\text{Edd} = 1.3 \times 10^{ 38} M_\text{BH} / M_\odot$ is the Eddington luminosity.

As for the angle $i$, its reliable determination from observational data is a rather complicated and not yet fully solved problem. Many authors often just assume $i \approx 40^{\circ} - 45^{\circ}$. In this work, we used the following approach. For each object, we took the average value of the angle $i$ from the literature, obtained by various methods \citep{inclination_sed, inclination_marin}. If the angle is unknown, the value $i = 45^{\circ}$ was taken, and if the current numerical method did not give a physically meaningful result, then the angles were alternately changed to the smaller and larger side with a step of 5$^\circ$ until having a result.

The spin value is determined numerically using the method from \citet{bardeen72}:

\begin{equation}
  \varepsilon(a) = 1 - \frac{R_\text{ISCO}^{3/2} - 2 R_\text{ISCO}^{1/2} + |a|}{R_\text{ISCO}^ {3/4}\left(R_\text{ISCO}^{3/2} - 3 R_\text{ISCO}^{1/2} + 2 |a|\right)^{1/2}},
  \label{eq01}
\end{equation}

\noindent where $R_\text{ISCO}$ is the radius of the innermost stable circular orbit, expressed in terms of spin as follows:

\begin{equation}
  \begin{array}{l}
   R_\text{ISCO}(a) = \\
   = 3 + Z_2 \pm [(3 - Z_1)(3 + Z_1 + 2 Z_2)]^{1/2},\\
   Z_1 = 1 + (1 - a^2)^{1/3}\left[(1 + a)^{1/3} + (1 - a)^{1/3}\right],\\
   Z_2 = (3 a^2 + Z_1^2)^{1/2}.
  \end{array}
  \label{eq02}
\end{equation}

\noindent Here "-"{} is used when $a \geq 0$, and "+"{} when $a < 0$.

\section{Spin evaluation}

To determine the values of spins $a$ and inclination angles $i$, it is necessary to know such AGN parameters as the SMBH mass $M_\text{BH}$ and the bolometric luminosity $L_\text{bol}$. These data were taken from the literature. In the case when only the luminosity at 5100\AA\, $L_{5100}$ could be found for an object, the bolometric luminosity was determined using the bolometric correction $BC$ as $L_\text{bol} = L_{5100} \times BC$. Determining the bolometric correction is quite a difficult task. In the literature, one can find very different values of $BC$ \citep{richards06, hopkins07, cheng19, netzer19, duras20}. In this paper, for consistency with our previous ones, we took the value $BC = 10.3$ \citep{richards06}, obtained from observations of more than 250 quasars and currently the most commonly used for estimating bolometric luminosity from optical data (in the rest frame), e.g., in new papers on \textit{JWST} data \citep{jwst}.

Since all spin determination models used by us assume the model of a geometrically thin, optically thick Shakura-Sunyaev \citep{shakura73} accretion disk, the objects should meet the condition $0.01 < l_\text{E} < 0.3$ \citep{netzer14}. 10 of the 14 objects from the sample fit the requirements. As for the remaining objects, we can assume geometrically thick accretion disks (ADAF) there.

Table \ref{tab02} presents the physical parameters of our objects taken from the literature and the values of the accretion disk inclination $i$ and the spin $a$ calculated on their basis for three models: (1) \citet{du14}, (2) \citet{raimundo11}, (3) \citet{trakhtenbrot14}. The sign ''---'' means that this method does not allow estimating the spin for this object.

One can note the large value of the angle $i \sim 65^{\circ}-75^{\circ}$ for Mrk~1018 estimated by all three models. This is somewhat larger than the result obtained by other methods ($\sim 45^{\circ}$), but this value lies within the possible limits concerning the object's type (Sy1.9).

In addition, we note the large value of the angle $i \sim 70^{\circ}-85^{\circ}$ for QSO~B0624+6907, which also exceeds the value obtained by other methods $\sim 45^{\circ }$ [note that in our work \citet{piotrovich17a} for this object, the value of the angle $i \approx 16^{\circ}$ was obtained by another method]. Such a large value of the angle may already indicate that the method used to determine the spin is not quite suitable for this object. In addition, the observed spectropolarimetric characteristics of this object show signs of equatorial scattering, indicating that we see effects from the interaction of radiation from the accretion disk and BLR with the dust torus.

\begin{table*}
\caption {Physical parameters of objects. $L_\text{bol}$ is the bolometric luminosity, $l_\text{E}$ is the Eddington ratio, $M_\text{BH}$ is the SMBH mass, $i_{1,2,3}$, $\varepsilon_{1,2,3}$ and $a_{1,2,3}$ is the angle between the line of sight and the normal to the accretion disk plane in degrees, radiative efficiency and the spin value for three models: (1) \citet{du14}, (2) \citet{raimundo11}, (3) \citet {trakhtenbrot14}, $i_\text{SED}$ - the average value of the angle $i$ from the literature, obtained for this object by other methods \citep{inclination_sed, inclination_marin}, if the angle is unknown, then the value of 45$^\circ$ was taken.} \label{tab02}
\begin{tabular}{lccccccccccccc}
\hline
Object & $L_\text{bol}$ & $\log{(l_\text{E})}$ & $\log{\left(\frac{M_\text{BH}}{M_\odot}\right)}$ & $i_1$ & $\varepsilon_1$ & $a_1$ & $i_2$ & $\varepsilon_2$ & $a_2$ & $i_3$ & $\varepsilon_3$ & $a_3$ & $i_\text{SED}$\\
\hline
PG 0052+251      & 45.82$^{2}$ & -0.75 & 8.46$^{2}$ & 40  & 0.301 & 0.998  & 20 & 0.166 & 0.922  & 20 & 0.288 & 0.996  & $\sim$20$^{8}$\\
LEDA 3095506     & 44.24$^{5}$ & -1.34 & 7.47$^{5}$ & 45  & 0.168 & 0.924  & 45 & 0.138 & 0.860  & 45 & 0.180 & 0.940  & $\sim$45$^{8}$\\
Mrk 359          & 43.55$^{3}$ & -0.79 & 6.23$^{3}$ & --- & ---   & ---    & 30 & 0.046 & -0.402 & 30 & 0.043 & -0.564 & $\sim$30$^{9}$\\
Mrk 1018         & 44.74$^{3}$ & -1.97 & 8.60$^{3}$ & 75  & 0.283 & 0.994  & 65 & 0.302 & 0.998  & 75 & 0.255 & 0.988  & $\sim$45\\
Mkn 590          & 44.12$^{3}$ & -1.56 & 7.57$^{7}$ & 35  & 0.304 & 0.998  & 25 & 0.297 & 0.996  & 40 & 0.309 & 0.998  & $\sim$25$^{9}$\\
LEDA 11567       & 43.81$^{5}$ & -0.88 & 6.58$^{5}$ & 40  & 0.040 & -0.754 & 45 & 0.045 & -0.452 & 45 & 0.046 & -0.402 & $\sim$45\\
QSO B0624+6907   & 46.10$^{6}$ & -1.61 & 9.60$^{6}$ & 85  & 0.116 & 0.778  & 70 & 0.259 & 0.990  & 80 & 0.220 & 0.974  & $\sim$45\\
LEDA 2196699     & 45.54$^{1}$ & -0.60 & 8.03$^{1}$ & 45  & 0.137 & 0.858  & 45 & 0.065 & 0.232  & 45 & 0.101 & 0.690  & $\sim$45$^{8}$\\
PG 1307+085      & 45.86$^{6}$ & -0.78 & 8.53$^{2}$ & 45  & 0.300 & 0.998  & 45 & 0.118 & 0.788  & 45 & 0.210 & 0.968  & $\sim$45\\
LEDA 3096673     & 44.87$^{4}$ & -0.97 & 7.73$^{4}$ & 40  & 0.167 & 0.922  & 40 & 0.102 & 0.696  & 40 & 0.145 & 0.880  & $\sim$40$^{8}$\\
\hline
\multicolumn{14}{l}{Data sources: (1) \citet{kim16}; (2) \citet{jha22}; (3) \citet{marin16}; (4) \citet{jin12a}; (5) \citet{sexton21}; (6) \citet{decarli08};}\\
\multicolumn{14}{l}{(7) \citet{bentz18}; (8) \citet{inclination_sed}; (9) \citet{inclination_marin}.}\\
\end{tabular}
\end{table*}

\begin{figure*}
\includegraphics[bb= 70 5 735 535, clip, width=0.66 \columnwidth]{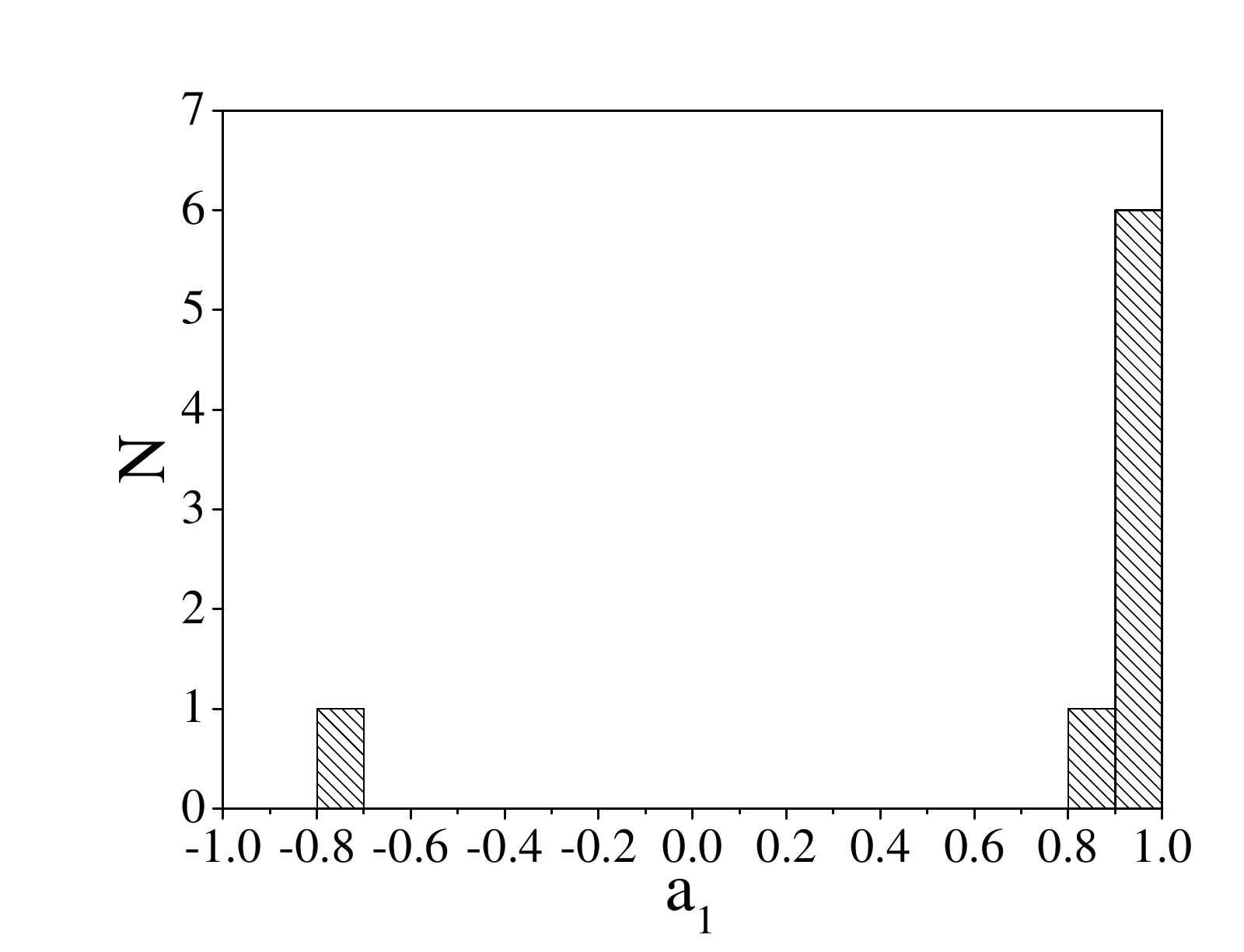}
\includegraphics[bb= 70 5 735 535, clip, width=0.66 \columnwidth]{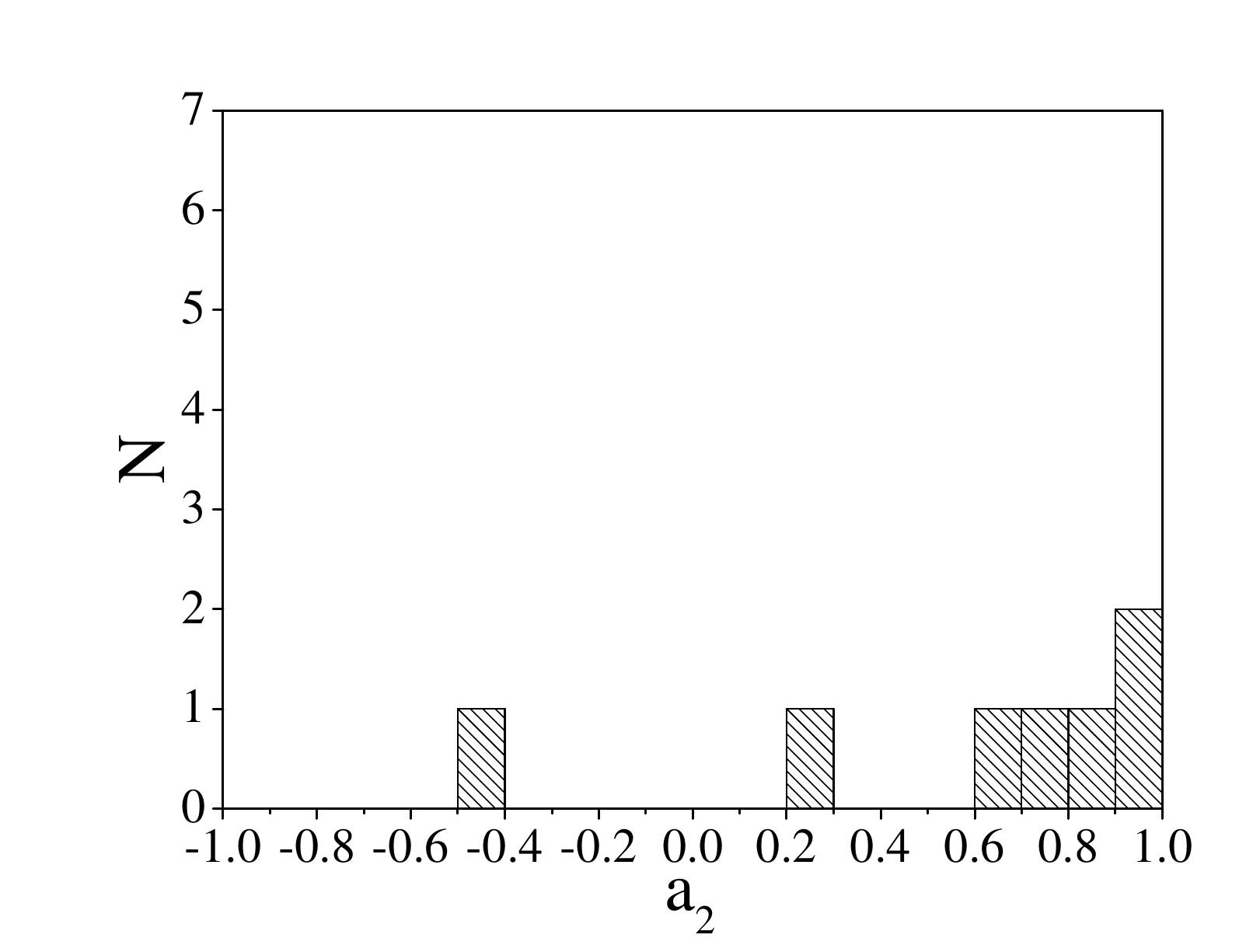}
\includegraphics[bb= 70 5 735 535, clip, width=0.66 \columnwidth]{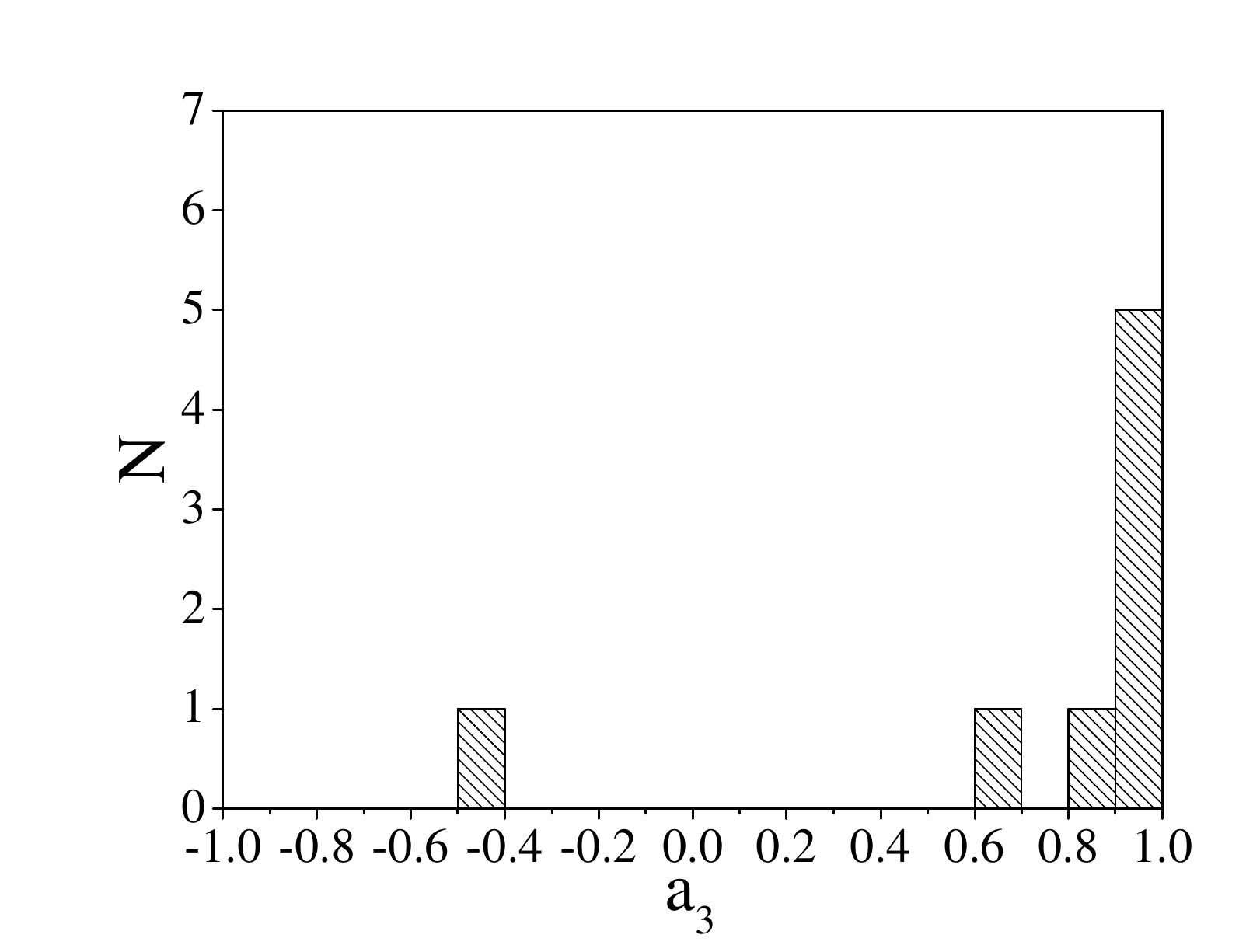}
\caption{Histograms showing the distribution of spin values for three models.}
  \label{fig01}
\end{figure*}

Figure \ref{fig01} shows histograms with the distribution of spin values for the three models. Obviously, due to the small number of objects, serious statistical analysis, in this case, is difficult, but it can be noted that the general qualitative form of distributions (especially for models 1 and 3), in which a pronounced peak appears for the values $0.9 < a < 1.0$, generally similar to distributions from our previous works \citep{afanasiev18,piotrovich22} and works of other authors \citep{trakhtenbrot14,daly19}.

\section{Determination of the characteristics of the magnetic field in the accretion disk}

When considering the magnetic field in an accretion disk, it is usually assumed \citep[see, e.g.,][]{pariev03} that its radius dependence has a power-law form:

\begin{equation}
  B(R) = B_\text{H} (R_\text{H} / R)^s,
  \label{eq03}
\end{equation}

\noindent where $B_\text{H}$ is the value of the magnetic field strength at the SMBH event horizon in AGN, $R_\text{H} = G M_\text{BH} \left(1 + \sqrt{1 - a^2 }\right) / c^2$ is the radius of the event horizon. As for the $s$ parameter, there are models with different values \citep{pariev03} of this parameter, but for the Shakura-Sunyaev disk, the value $s = 1.25$ \citep{shakura73} is most often taken.

To determine the parameters of the magnetic field in the accretion disk, we used a numerical simulation method developed by us on the basis of \citep{silantev09}. It is described in detail in our article \citet{piotrovich21}. The method is based on the calculation of the wavelength-dependent integral polarization of radiation from an accretion disk in the presence of a magnetic field, which makes it possible to estimate the field geometry from the characteristics of the observed polarization. Since the dependence of the polarization on the wavelength in this model turns out to be rather weak compared to the error of spectropolarimetric observations, in this case, we used polarization values averaged over the observed range.

Since here we consider only polarized radiation from the accretion disk itself, we had to exclude from further consideration QSO~B0624+6907, which, as noted above, shows signs of equatorial scattering and has additional mechanisms of polarization. In addition, we note that signs of equatorial scattering are also observed in the object LEDA~3095636, for which at the previous stage it was not possible to determine the spin due to the values of the Eddington ratio that are unsuitable for the models used.

Table \ref{tab03} presents the observed polarization and the calculated values of the magnetic field at the SMBH event horizon $B_\text{H}$ and the power law parameter of the magnetic field on the radius $s$ for the three models used. Note that for the object Mrk~359, the model used did not give a result for any of the values of the spin and angle.

\renewcommand{\arraystretch}{1.5}
\begin{table*}
\caption {The calculated values of the magnetic field strength at the SMBH event horizon $B_\text{H,1,2,3}$ and the parameter of the power-law dependence of the magnetic field on the radius $s_{1,2,3}$ for three models: (1) \citet{du14}, (2) \citet{raimundo11}, (3) \citet{trakhtenbrot14}, $P$ - observed polarization in \%, magnetic field in Gauss.} \label{tab03}
\begin{tabular}{lccccccc}
\hline
Object & $P$ & $\log{\left(B_\text{H,1} [G]\right)}$ & $s_1$ & $\log{\left(B_\text{H,2} [G]\right)}$ & $s_2$ & $\log{\left(B_\text{H,3} [G]\right)}$ & $s_3$ \\
\hline
PG 0052+251  & 0.44$\pm$0.34 & $4.29^{+0.30}_{-1.29}$ & 1.32$\pm$0.36 & $3.80^{+0.30}_{-0.80}$ & 1.66$\pm$0.26 & $3.89^{+0.30}_{-0.89}$ & 1.65$\pm$0.26\\
LEDA 3095506 & 2.45$\pm$0.71 & $3.59^{+0.28}_{-0.59}$ & 1.76$\pm$0.19 & $3.56^{+0.27}_{-0.56}$ & 1.77$\pm$0.18 & $3.61^{+0.28}_{-0.61}$ & 1.76$\pm$0.19\\
Mrk 1018     & 1.62$\pm$0.30 & $3.98^{+0.29}_{-0.98}$ & 1.40$\pm$0.36 & $3.82^{+0.26}_{-0.78}$ & 1.49$\pm$0.31 & $3.96^{+0.29}_{-0.96}$ & 1.39$\pm$0.36\\
Mkn 590      & 1.25$\pm$0.29 & $3.56^{+0.27}_{-0.56}$ & 1.77$\pm$0.18 & ---                    & ---           & $3.85^{+0.28}_{-0.85}$ & 1.63$\pm$0.23\\
LEDA 11567   & 2.58$\pm$0.85 & $3.77^{+0.30}_{-0.77}$ & 1.77$\pm$0.18 & $4.03^{+0.30}_{-1.03}$ & 1.68$\pm$0.24 & $4.03^{+0.30}_{-1.03}$ & 1.69$\pm$0.24\\
LEDA 2196699 & 0.48$\pm$0.36 & $4.33^{+0.30}_{-1.33}$ & 1.21$\pm$0.36 & $4.31^{+0.30}_{-1.31}$ & 1.26$\pm$0.37 & $4.32^{+0.30}_{-1.32}$ & 1.24$\pm$0.36\\
PG 1307+085  & 1.54$\pm$0.66 & $3.84^{+0.30}_{-0.84}$ & 1.64$\pm$0.24 & $3.66^{+0.28}_{-0.66}$ & 1.69$\pm$0.22 & $3.77^{+0.30}_{-0.77}$ & 1.66$\pm$0.23\\
LEDA 3096673 & 2.01$\pm$0.33 & $3.40^{+0.21}_{-0.40}$ & 1.84$\pm$0.13 & $3.34^{+0.18}_{-0.31}$ & 1.86$\pm$0.12 & $3.38^{+0.20}_{-0.38}$ & 1.84$\pm$0.13\\
\hline
\end{tabular}
\end{table*}
\renewcommand{\arraystretch}{1.0}

\begin{figure*}
\includegraphics[bb= 70 0 735 535, clip, width=0.66 \columnwidth]{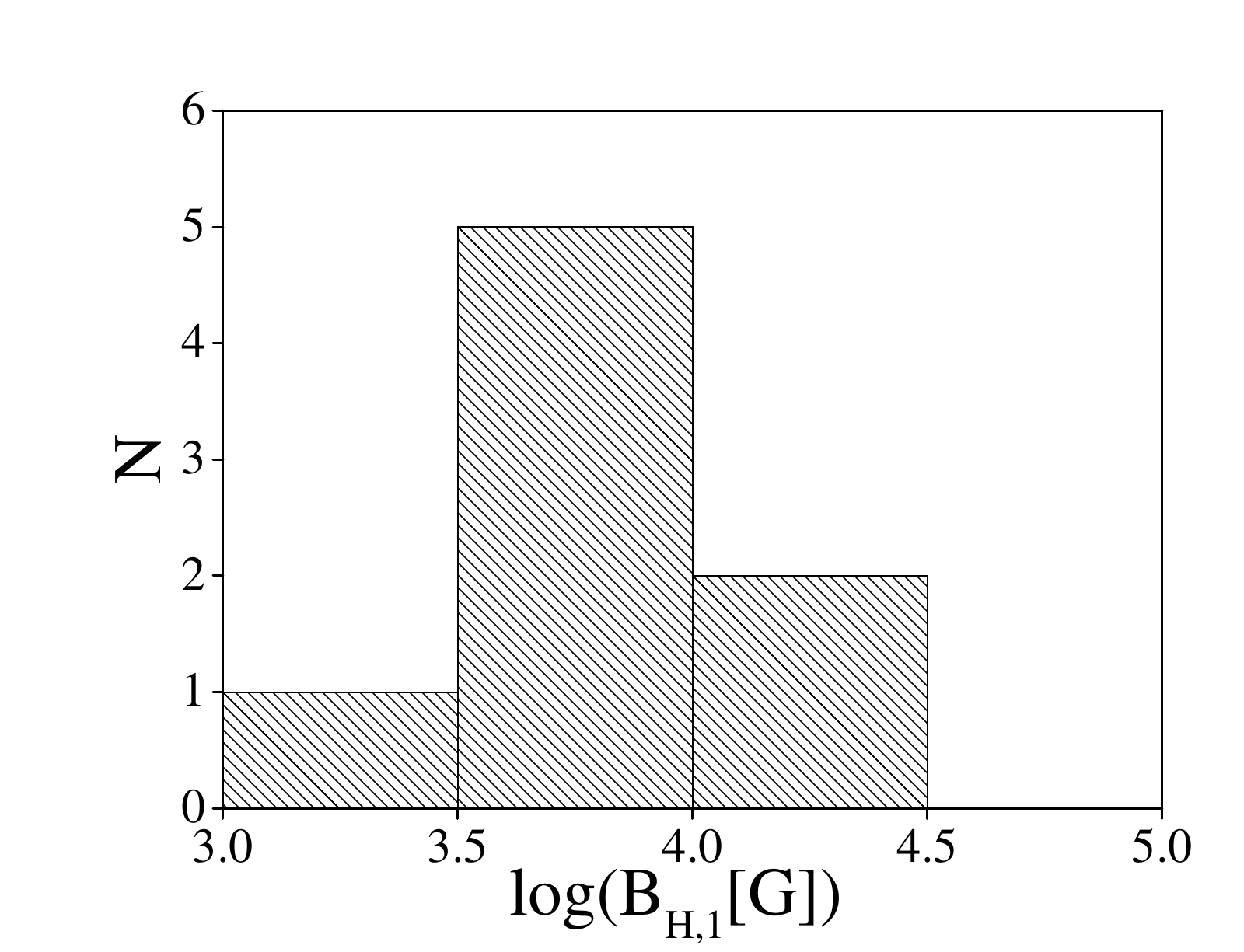}
\includegraphics[bb= 70 0 735 535, clip, width=0.66 \columnwidth]{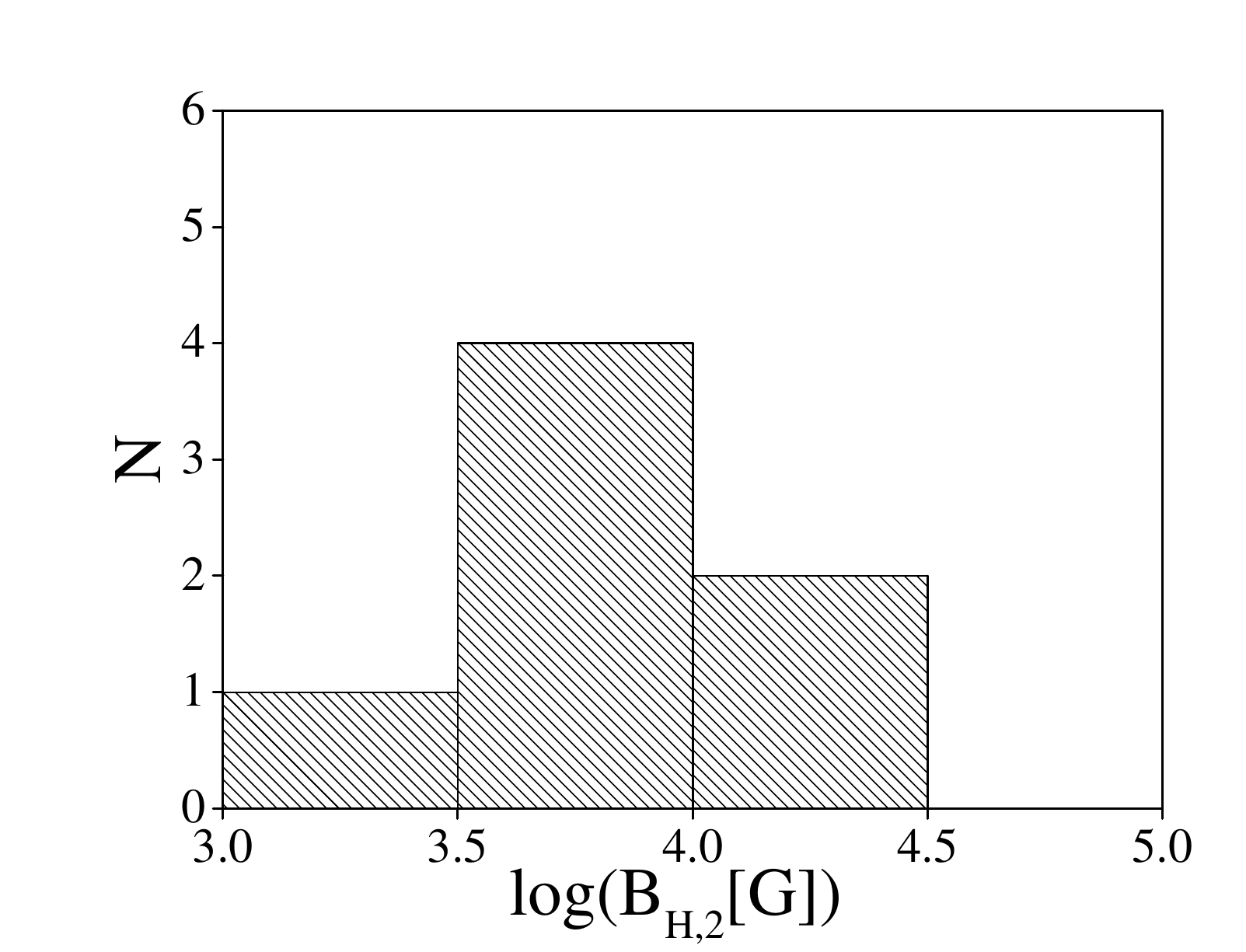}
\includegraphics[bb= 70 0 735 535, clip, width=0.66 \columnwidth]{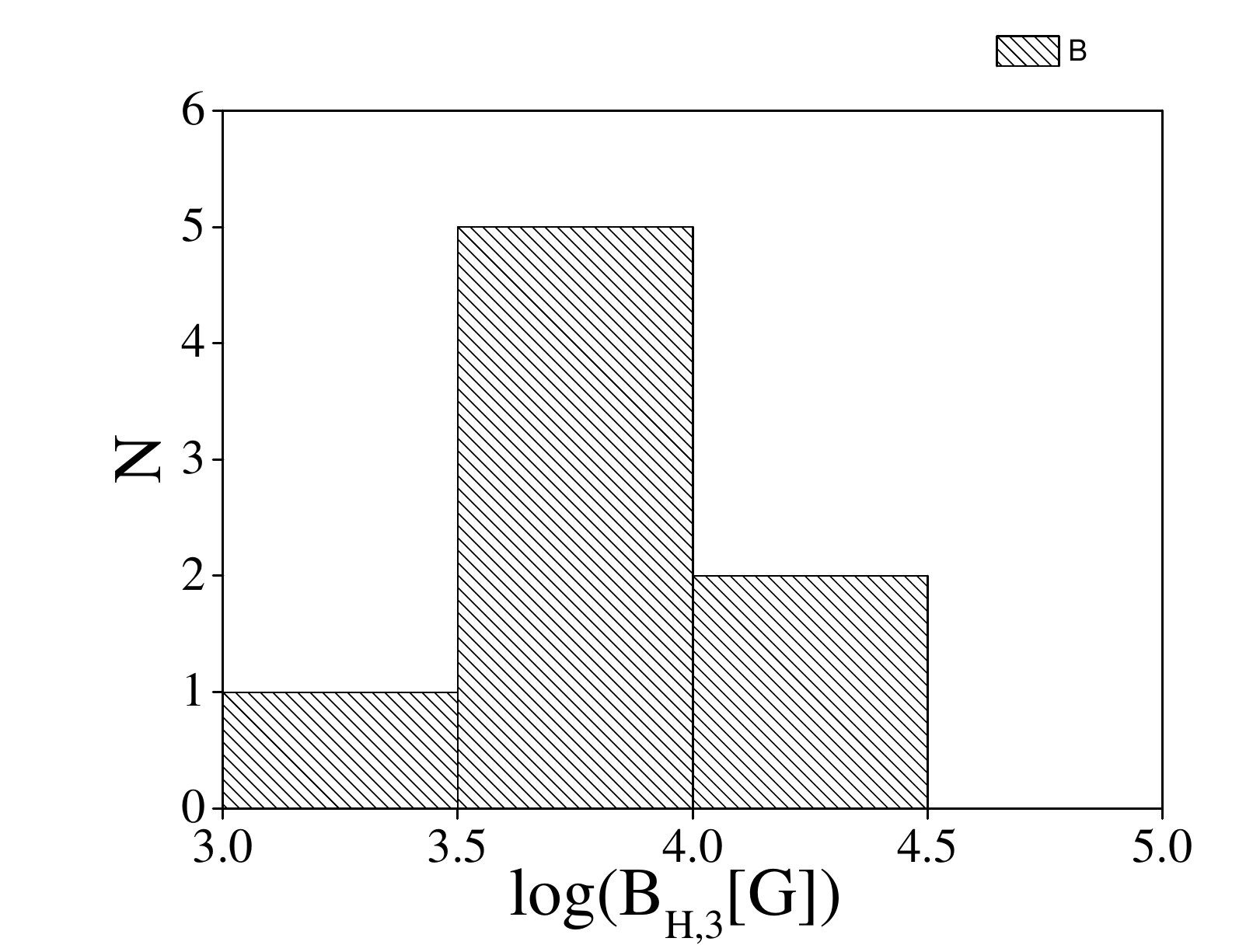}
\caption{Histograms showing the distribution of magnetic field values at the SMBH event horizon for three models.}
 \label{fig02}
\end{figure*}

\begin{figure*}
\includegraphics[bb= 70 5 735 535, clip, width=0.66 \columnwidth]{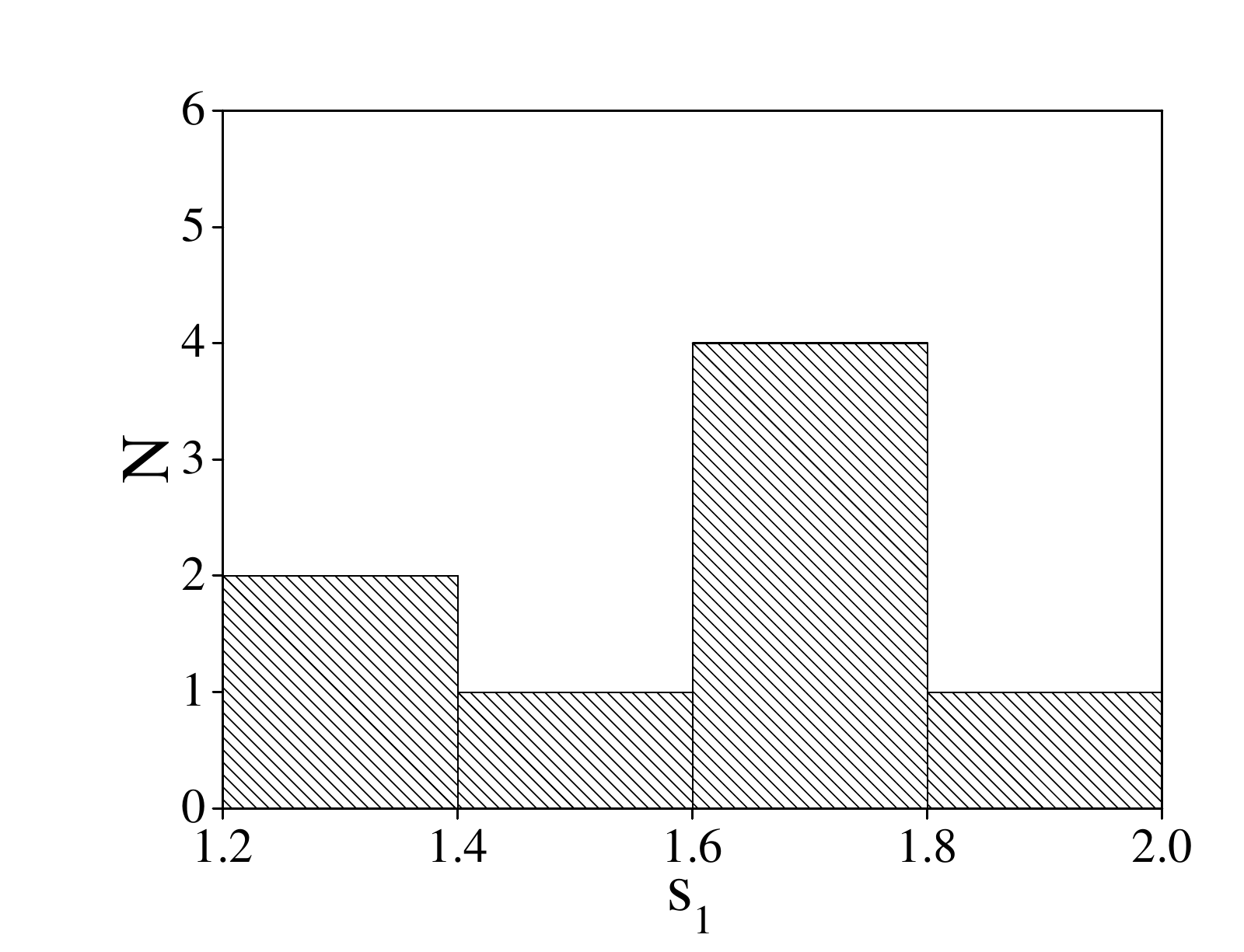}
\includegraphics[bb= 70 5 735 535, clip, width=0.66 \columnwidth]{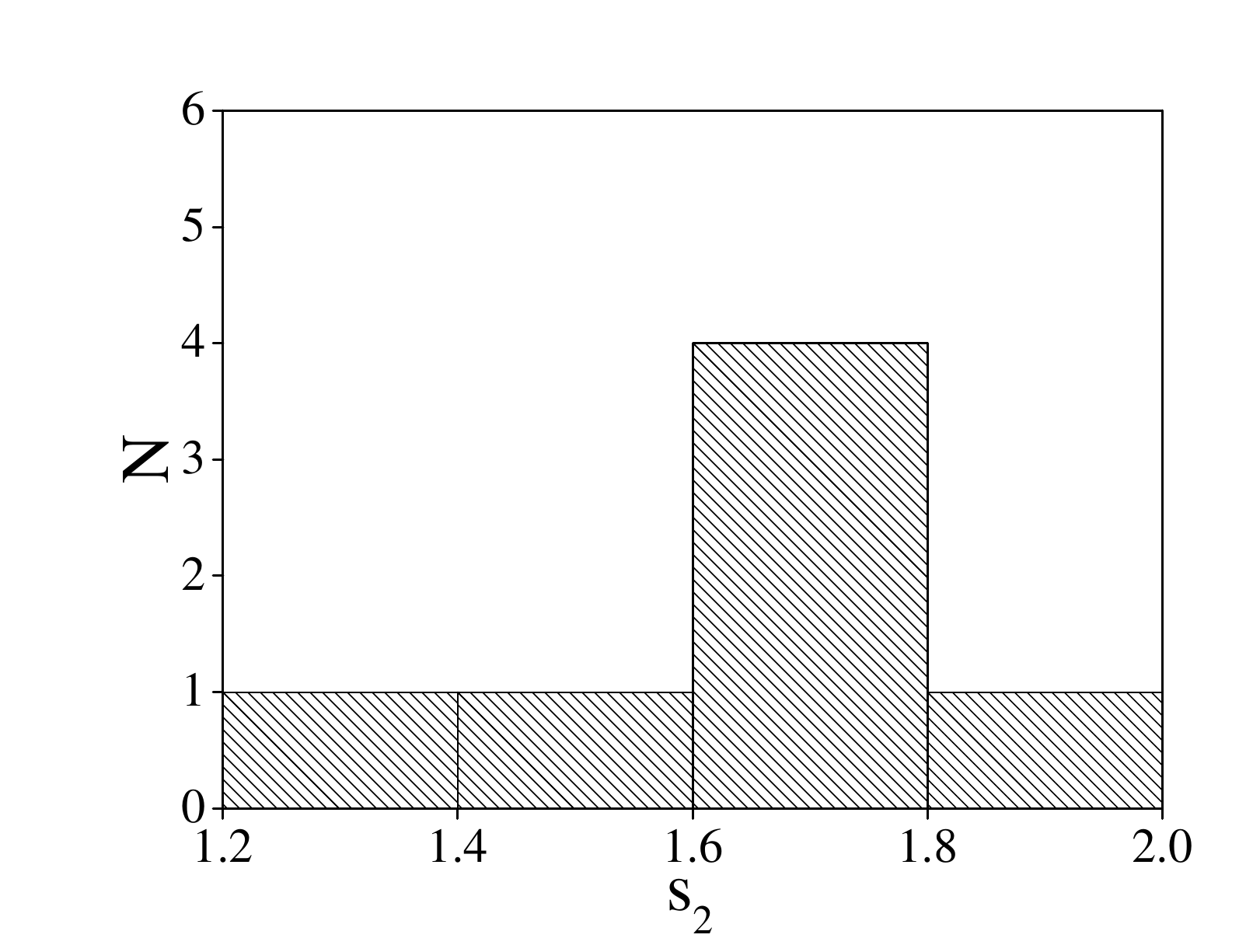}
\includegraphics[bb= 70 5 735 535, clip, width=0.66 \columnwidth]{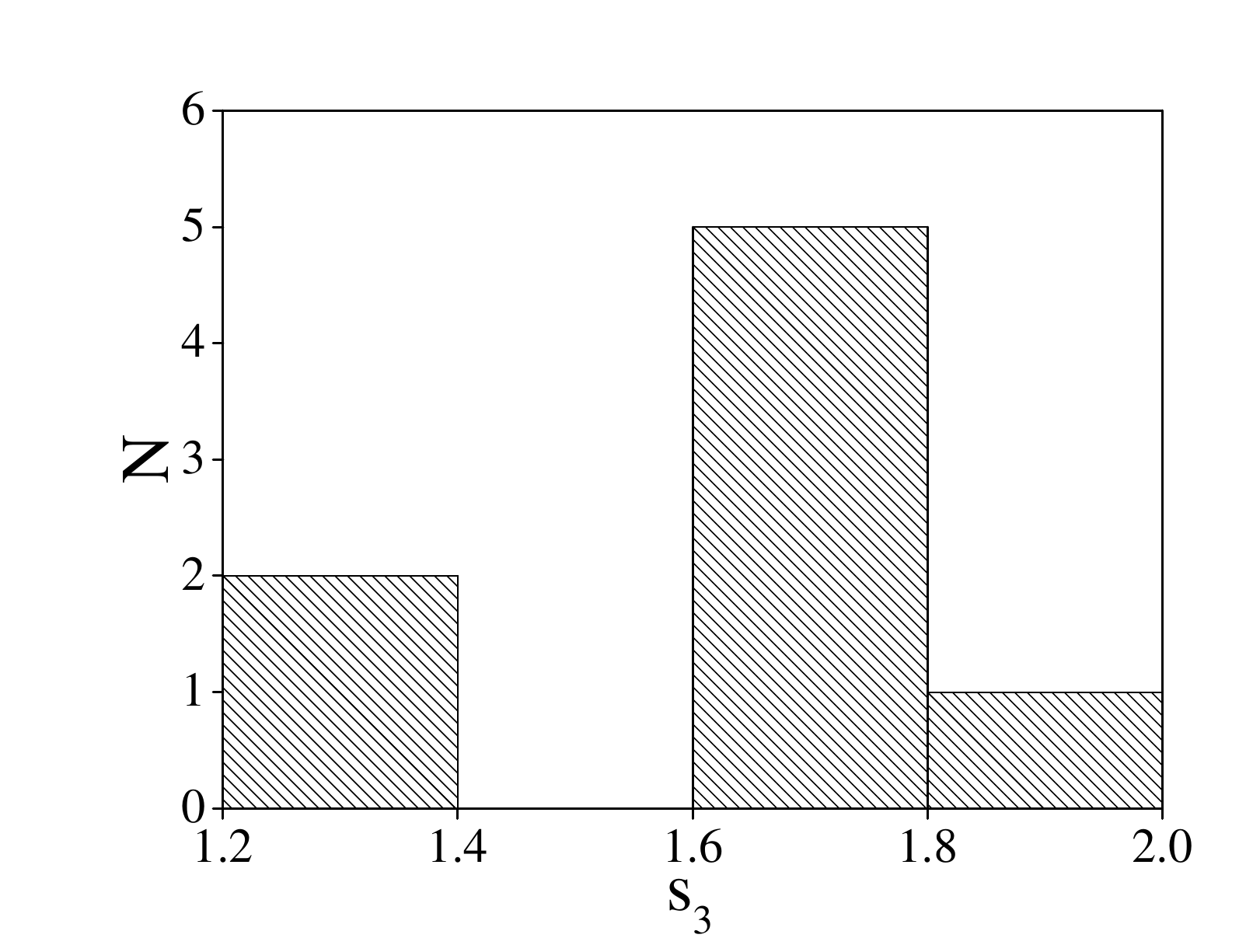}
\caption{Histograms showing the distribution of the values of the parameter of the power-law dependence of the magnetic field on the radius for three models.}
 \label{fig03}
\end{figure*}

As mentioned earlier, a small number of objects does not allow for serious statistical analysis, but we decided to at least qualitatively evaluate the distributions of the obtained parameters.

In Fig. \ref{fig02} histograms with the distribution of magnetic field values at the event horizon for three models are presented. The shape of the distributions qualitatively looks similar for all models. All models show a peak around $3.5 < \log{(B_\text{H})} < 4.0$. This result agrees well with the results from our paper \citet{piotrovich20} and from \citet{daly19}, in which the magnetic field was estimated by different methods. Note that in our work \citet{piotrovich21} the distribution was also similar, but the peak was in the range $4.0 < \log{(B_\text{H})} < 4.5$, which can be explained by the insufficient number of objects and the selection effect.

It should be noted that in \citet{daly19} values of spin and magnetic field are estimated for 3 objects from our sample: PG~0052+251, Mkn~590 and PG~1307+085. The magnetic field strength obtained in \citet{daly19} for PG~0052+251 is $\log{B[G]} = 4.49 \pm 0.37$. This is consistent with our values: $4.29_{+0.30}^{-1.29}$, $3.80_+{0.30}^{-0.80}$ and $3.89_{+0.30}^{-0.89}$. All three of our values agree to better than 1$\sigma$ with the value obtained with the completely independent ''outflow method'' in \citet{daly19}. For PG~1307+085, the value obtained by Daly is $4.39 \pm 0.37$. This is consistent with our values: $3.84_{+0.30}^{-0.84}$, $3.66_{+0.28}^{-0.66}$ and $3.77_{+0.30}^{-0.77}$ at about or better than 1$\sigma$. For Mkn~590, the value obtained by Daly is $5.05 \pm 0.37$. This is consistent with two of our values: $3.56_{+0.27}^{-0.56}$ and $3.85_{+0.28}^{-0.85}$ at about the 2.3$\sigma$ level.

For the spin values, \citet{daly19} obtained a value of $0.57 \pm 0.15$ for PG~0052+251, $0.41 \pm 0.15$ for PG~1307+085 and $0.88 \pm 0.15$ for Mkn~590. We obtain the values between about 0.9 and 1.0 for PG~0052+251; between about 0.8 and 1.0 for PG~1307+085; and values close to one for Mkn~590. The spin values obtained with the independent methods: agree for Mkn~590 at better than 1$\sigma$; and agree at 1 to 2 $\sigma$ for PG~0052+251 and PG~1307+085. Thus, both the magnetic field strengths and the black hole spin values obtained with our continuum fitting method agree to within 1 to 2 $\sigma$ with the values obtained with the outflow method by Daly for three sources for which a comparison is possible. Also, the statistical properties of our sample of objects are quite close to the properties of the sample from \citet{daly19}.

Also in \citet{daly21} the radiative efficiency factors were obtained for the same three objects by ''outflow method'' which is completely independent of that used by us. The efficiency factors obtained for PG~0052+251, PG~1307+085, and Mkn~590 are $0.48 \pm 0.35$, $0.34 \pm 0.25$ and $0.49 \pm 0.36$, respectively. The values obtained by \citet{daly21} are consistent with those obtained with our method (see Table \ref{tab02}).

In Fig. \ref{fig03} distributions of parameter $s$ for three models are presented. It can be seen that the shape of the distribution for all models looks similar in general terms, and in all cases, there is a distinct peak in the range of $1.6 < s < 1.8$. This type of distribution is somewhat different from the similar distribution from our paper \citet{piotrovich21}, in which the peak fell in the range $1.85 < s < 2.0$. Most likely, this is also due to the small number of objects and the selection effect. However, it should be noted that in both cases the peak in the distribution of the $s$ parameter indicates a significantly larger value than the standard value from the Shakura-Sunyaev model $s = 1.25$ \citep{shakura73}, and most of the objects have the value $s > 1.25$. This may indicate that the classical Shakura-Sunyaev model is not suitable for accurately describing accretion disks in objects of this type, and some modification may be required.

\section{Conclusions}

Based on the data of spectropolarimetric observations of 14 active galaxies at the BTA-6m telescope of the SAO RAS, as well as on the basis of literature data, we made the estimates of the AGN physical parameters: the value of the spin of the central SMBH, the angle between the line of sight and the normal to the accretion disk plane (appeared to be close to the estimates obtained by independent methods), the magnetic field strength at the SMBH event horizon, and the value of the parameter of the power-law dependence of the magnetic field strength on the radius in the accretion disk.
The advantages of the work include the fact that using high-quality spectropolarimetric data, we studied the AGN sample that is less distorted by third-party mechanisms for the occurrence of polarization.

The spin distribution has a pronounced peak in the region of $0.9 < a < 1.0$. This result is in good agreement with our previous works \citep{afanasiev18,piotrovich22} and the works of other authors \citep{trakhtenbrot14}. The distribution of magnetic SMBH values at the event horizon exhibits a peak in the region of $3.5 < \log{(B_\text{H})} < 4.0$. This result also agrees well with the result from our paper \citet{piotrovich20}, in which the magnetic field was estimated by a different method. However, it differs somewhat from the result from \citet{piotrovich21}, in which the peak was around $4.0 < \log{(B_\text{H})} < 4.5$, which is most likely due to low statistics and the selection effect.

In the distribution of the parameter $s$ of the power-law dependence of the magnetic field on the radius, there is a distinct peak in the region of $1.6 < s < 1.8$, which is somewhat different from the result from our paper \citet{piotrovich21}, in which the peak was at $1.85 < s < 2.0$. Apparently, this is due to the small number of objects and the selection effect. However, an important result is the fact that in both cases the peak in the distribution of the $s$ parameter falls on a value significantly larger than the standard value from the Shakura-Sunyaev model $s = 1.25$ \citep{shakura73}, and most of the objects have the value $s > 1.25$. This may indicate that the classical Shakura-Sunyaev model needs to be modified to more accurately describe accretion disks in objects of this type.

\section*{Acknowledgements}

This work is a continuation of the cycle of works, the foundation of which was laid by Professor V.L. Afanasiev (instrumental support) and Professor Yu.N. Gnedin (theoretical part). The authors are grateful for the support during observations on the BTA-6m to the staff of the Special Astrophysical Observatory: A.V. Moiseev and R.I. Uklein.

M.Yu. Piotrovich, S.D. Buliga and T.M. Natsvlishvili were supported by the state order of the Central Astronomical Observatory at Pulkovo, the planned research topic "MAGION"{} - Physics and evolution of stars and active galactic nuclei.

E.S. Shablovinskaya and E.A. Malygin obtained observed data on the unique scientific facility "Big Telescope Alt-azimuthal" of SAO RAS as well as made data processing with the financial support of grant No075-15-2022-262 (13.MNPMU.21.0003) of the Ministry of Science and Higher Education of the Russian Federation. E.S. Shablovinskaya acknowledges support from ANID BASAL project FB210003 and Gemini ANID ASTRO21-0003.

Observations with the SAO RAS telescopes are supported by the Ministry of Science and Higher Education of the Russian Federation. The renovation of telescope equipment is currently provided within the national project "Science and Universities".

\section*{Data Availability}

The data underlying this article are available in the article.

\bibliographystyle{mnras}
\bibliography{mybibfile} 

\bsp	
\label{lastpage}
\end{document}